\documentclass[runningheads]{llncs}
\usepackage[T1]{fontenc}
\usepackage{graphicx}
\usepackage{bm}
\begin{document}
\title{Wave packet treatment of neutrino flavour and spin oscillations in galactic and extragalactic magnetic fields}
\titlerunning{Wave packet treatment of neutrino of neutrino oscillations}
\author{Artem Popov  \and
Alexander Studenikin }
\authorrunning{A. Popov et al.}
\institute{Department of Theoretical Physics, \\ Moscow State University, 119991 Moscow, Russia \\
\email{ar.popov@physics.msu.ru, studenik@srd.sinp.msu.ru}}
\maketitle              % typeset the header of the contribution
\begin{abstract}
We consider neutrino flavour and spin oscillations in a magnetic field using formalism of wave packets. Decoherence effects due to neutrino wave packets separation are studied. The considered effects are especially important for describing astrophysical neutrino oscillations, since they propagate on kiloparsec scale and bigger. The obtained results are of interest for neutrino telescopes IceCube, Baikal-GVD and KM3NeT, and also can be applied for description of supernovae neutrino oscillations effects would be detected by JUNO and Hyper-Kamiokande.

\keywords{Neutrino magnetic moment  \and Neutrino oscillations \and Astrophysics.}
\end{abstract}
\section{Introduction}
It is known that massive neutrinos possess nontrivial electromagnetic properties, in particular nonzero anomalous magnetic moments. For a review of theory and experiment of neutrino electromagnetic properties see \cite{Giunti:2014ixa} and \cite{Studenikin:2022rhv}. Currently, the best upper bound on neutrino effective magnetic moment obtained by the XENONnT experiment and is on the level $\mu_\nu \sim 6.4\times10^{-12}\mu_B$ \cite{XENONCollaboration:2022kmb,Giunti:2023yha}. Interaction of neutrino magnetic moments with a magnetic field induces neutrino spin and spin-flavour precession, i.e. a phenomena in which neutrinos are converted to right-handed states. In particular, spin precession can be induced by neutrino interaction with cosmic magnetic field, that is of order of $\mu$G in our Galaxy \cite{Beck:2008ty}.

To study neutrinos propagation through an environment of cosmological scales it is necessary to employ the wave packet approach that accounts for decoherence of neutrino oscillations. Previously, the wave packet formalism was developed for the case of vacuum neutrino oscillations within relativistic quantum mechanics (see \cite{Giunti:2003ax} for a review) and quantum field theory \cite{Naumov:2020yyv} approaches, for neutrino oscillations in matter \cite{Peltoniemi:2000nw,Kersten:2015kio} and collective neutrino oscillations \cite{Akhmedov:2017mcc}. In this paper we extend the wave packet approach to the case of neutrino propagation in a magnetic field.

\section{Neutrino oscillations in a magnetic field in the wave packet formalism}
Neutrino oscillations in a uniform magnetic field are described by the following modified Dirac equation
\begin{equation}\label{Dirac_eq}
	(i\gamma^{\mu}\partial_{\mu} - m_i)\nu_i(x) - \mu_{i}\bm{\Sigma}\bm{B} \nu_i(x) = 0,
\end{equation}
where $i=1,2,3$ enumerates neutrino massive states. In \cite{Popov:2019nkr,Lichkunov:2022mjf}, equation (\ref{Dirac_eq}) was solved in the plane wave approximation. To account for potentially important decoherence effects in neutrino oscillations at long distances, we solve (\ref{Dirac_eq}) assuming that at initial moment $t=0$ neutrino wave function Fourier transform is described by a Gaussian wave packet
\begin{equation}
	\nu_i(p,0) = f_i(p,p_0) u^{-}_i(p), \;\;\;
	f_i(p,p_0) = \frac{1}{(2\pi\sigma_p^2)^{1/4}}\exp\left(-\frac{(p-p_0)^2}{4\sigma_p^2}\right),
\end{equation}
where $p_0$ is the average wave packet momentum, $\sigma_p$ is wave packet width in momentum space and $u^{-}_i$ is a left-handed solution of the Dirac equation in vacuum. For clarity, in this paper we consider one dimensional wave packets. Three-dimensional consideration leads to different wave packet spreading times in longitudinal and transversal directions, but does not significantly affect neutrino oscillations patterns \cite{Naumov:2020yyv,Naumov:2013uia}.

The dispersion relation for neutrinos interacting with a magnetic field is given by the following expression \cite{Popov:2019nkr}
\begin{equation}\label{energy}
	E_i^s(p) = \pm\sqrt{m_i^2 + p^2 + \mu_i^2 B^2 + 2s\mu_i\sqrt{m_i^2 B^2 + p^2 B_{\perp}^2 }},
\end{equation}
where $s = \pm 1$. Here magnetic field $\bm{B}$ is decomposed into transverse $\bm{B}_{\perp}$ and longitudinal $\bm{B}_{\parallel}$ components with respect to the neutrino momentum.

For sufficiently small $\sigma_p$, (\ref{energy}) can be decomposed near average momentum $p_0$
\begin{equation}\label{dispersion_decomposition}
	E_i^s(p) = E(p_0) + v_i^s(p_0)(p-p_0) + \mathcal{O}((p-p_0)^2),
\end{equation}
where $\frac{\partial E_i^s(p)}{\partial p}\Big|_{p=p_0}$ are group velocities. Using (\ref{energy}), we calculate neutrino wave packets group velocities accounting for neutrino interaction with a magnetic field
\begin{equation}\label{group_velocity}
	v_i^s(p_0) = \frac{p_0}{E_i^s(p_0)}\left(1+\frac{s\mu_i B^2_{\perp}}{\sqrt{m_i^2 B^2 + p^2_0 B^2_{\perp}}} \right).
\end{equation}

Finally, for the probabilities of neutrino flavour and spin oscillations in a magnetic field accounting for decoherence effects we get 
\begin{equation}\label{prob_2}
	P_{\nu_{\alpha}^h \to \nu_{\beta}^{h'}}(L) = \sum_{i,j} \sum_{s,s'} U^*_{\beta i} U_{\alpha i} U_{\beta j} U^*_{\alpha j} C_{is}^{hh'} C_{js'}^{hh'} \exp\Big( -i2\pi \frac{L}{L_{osc}^{ij ss'}}\Big) \exp\Big( -\frac{L^2}{ (L_{coh}^{ij ss'})^2} \Big),
\end{equation}
where the corresponding oscillations and coherence lengths are introduced

\begin{equation}
	\label{osc_coh_lengths}
	L_{osc}^{ij ss'} = \frac{\pi}{\omega_{ij}^{ss'}},  \;\;\;
	L_{coh}^{ij ss'} = \frac{2\sqrt{2}\sigma_x}{v_i^s-v_j^{s'}}.
\end{equation}
The oscillations frequencies $\omega_{ij}^{s\sigma}$ are given by $	\omega_{ij}^{s\sigma}(p_0)\approx \frac{\Delta m^2_{ij}}{2p_0} + (\mu_i s - \mu_j \sigma)B_{\perp}$.

The coefficients $C^{hh'}_{is}$ in (\ref{prob_2}) are given by
\begin{equation}\label{LL}
	C^{LL}_{i s} \approx \frac{1}{2} + \mathcal{O}\Big(\frac{m_i^2}{p^2}\Big), \;\;\; C^{RL}_{i s} \approx - \frac{s}{2} + \mathcal{O}\Big(\frac{m_i^2}{p^2}\Big).
\end{equation}

The oscillations probabilities (\ref{prob_2}) generalize the expressions we obtained in \cite{Lichkunov:2022mjf} and account for exponential damping of neutrino oscillations at large distance due to wave packets separation. Using (\ref{group_velocity}), we obtain the following approximate expressions for the coherence lengths given that $p \gg m_i \gg \mu_i B$:

\begin{equation}
	L_{coh}^{ijss} \approx  \frac{4\sqrt{2}\sigma_x p^2}{\Delta m_{ij}^2},\;\;\;
	L_{coh}^{ii-+}  \sim \frac{\sigma_x p^3}{\mu_iBm_i^2},\;\;\;
	L_{coh}^{ij-+} \approx L_{coh}^{ijss}.
\end{equation}
Here the coherence lengths $L_{coh}^{ijss}$ and $L_{coh}^{ii-+}$ describe damping of oscillations on vacuum lengths $L_{osc}^{ijss} = \frac{4 \pi p}{\Delta m^2_{ij}}$ and magnetic lengths $	L_{osc}^{ii-+} = \frac{\pi}{\mu_i B_{\perp}}$, correspondingly. Note that unlike the coherence lengths of vacuum neutrino oscillations, the coherence length $L_{coh}^{ii-+}$ is proportional to the cube of the neutrino momentum.

Interaction with a magnetic field can modify the flavour composition of neutrino emanating from astrophysical objects. For example,  consider high-energy neutrinos originating from the Galactic centre. The flavour composition at distance $L$ is given by
\begin{equation}\label{flavour_ratios}
	r_\alpha(L) = \frac{\sum_\alpha r^0_\alpha P_{\alpha \beta}(L)}{\sum_{\alpha\beta} r^0_\alpha P_{\alpha \beta}(L)},
\end{equation}
where $r_\beta^0$ are the flavour ratios at the source and $P_{\alpha \beta} = P_{\nu_{\alpha}^L \to \nu_{\beta}^{L}}$ are the probabilities of flavour oscillations.

\begin{figure}[tbp]
	\centering % \begin{center}/\end{center} takes some additional vertical space
	\includegraphics[width=.49\textwidth]{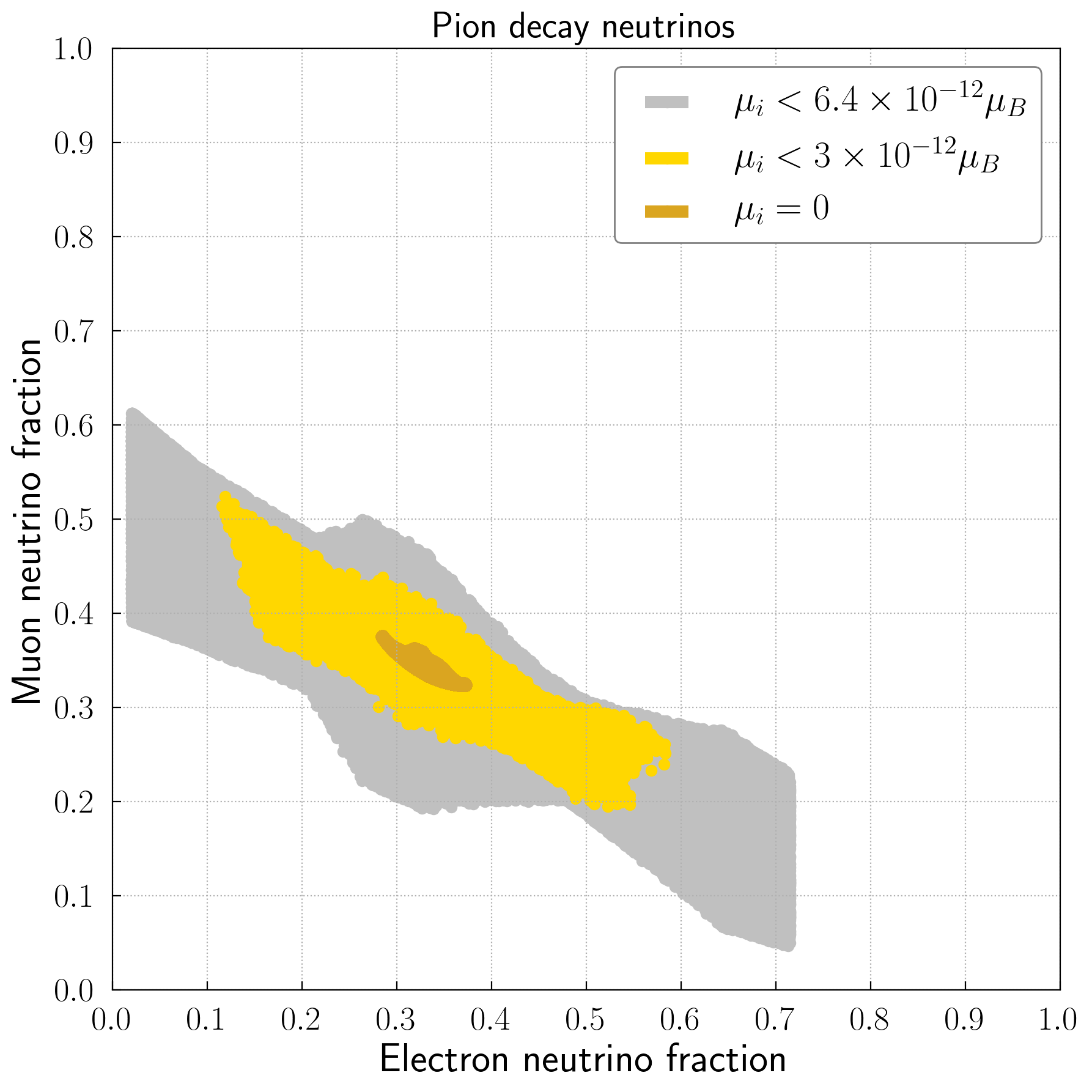}
	%	\hfill
	\includegraphics[width=.49\textwidth]{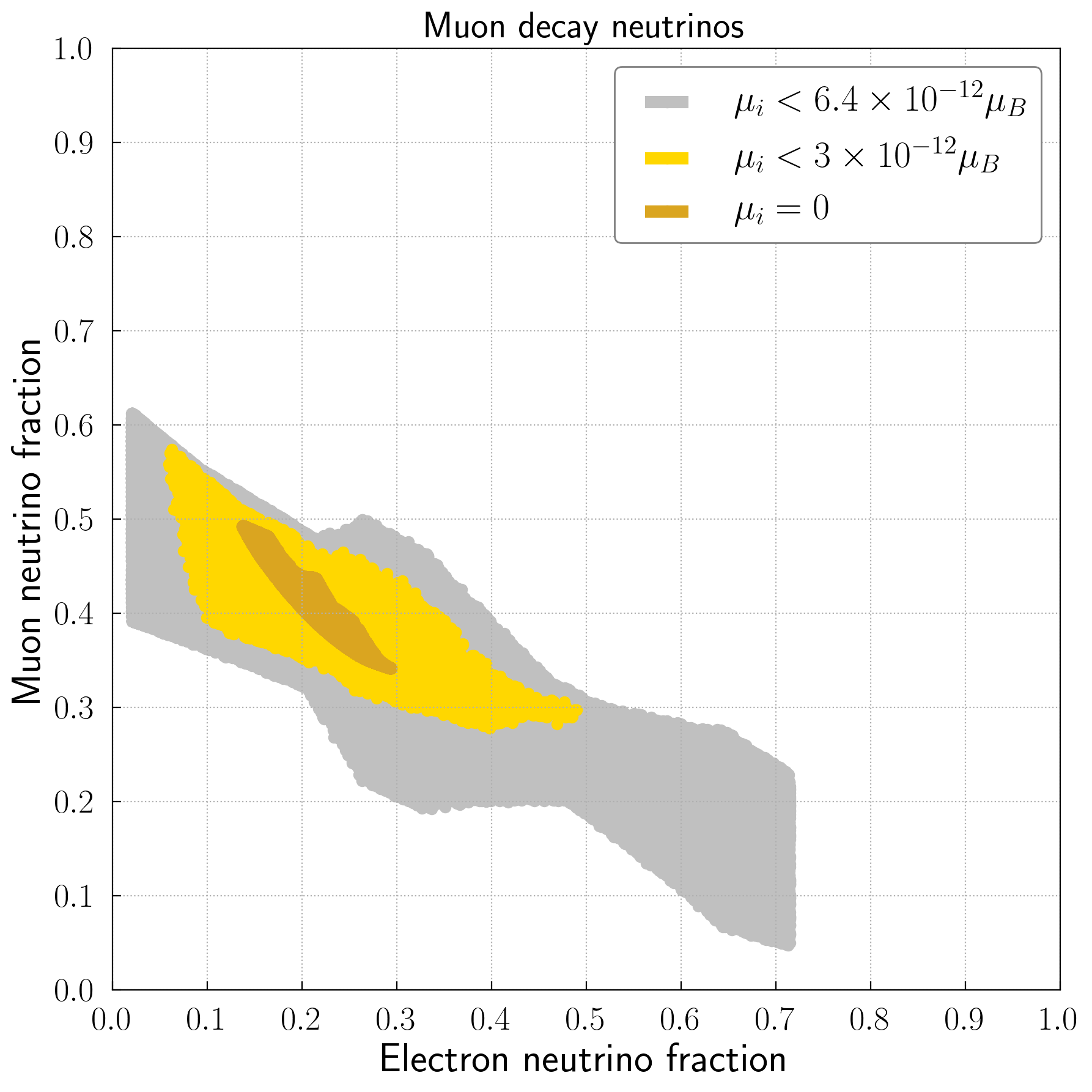}
	% "\includegraphics" is very powerful; the graphicx package is already loaded
	\caption{\label{fig:i} Flavour compositions of neutrinos coming from the Galactic centre for the cases of pion decay and muon decay high-energy neutrinos production.}
\end{figure}

The mechanism of high-energy neutrinos production is presently unknown. We study two cases considered in the literature: case of pion decay neutrinos and case of muon decay neutrinos. The corresponding initial flavour compositions are $r^0 = (\frac 1 3, \frac 2 3,0)$ and $r^0 = (0,1,0)$. We also consider different values of neutrino magnetic moments from the interval $(0.1, 6.4)\times 10^{-12}\mu_B$, as well as different experimental values of neutrino mixing parameters within their $3\sigma$ intervals. In Fig. \ref{fig:i} we show the possible flavour ratios of neutrinos propagating from the Galactic centre to a terrestrial neutrino telescope for different values of neutrino magnetic moments $\mu_1,\mu_2,\mu_3$ and compare them to ones predicted by the vacuum neutrino oscillations case. We assume that galactic magnetic field strength is $B = 0.2\;\mu$G, which corresponds to the average value of the regular component of the Galactic magnetic field. Magnetic field of approximately same strength is found in the Local Supercluster \cite{Valee:2002}. We find that the neutrino interaction with the Galactic magnetic field can significantly modify the observed flavour ratio. Thus, we conclude that neutrino magnetic moments potentially can be probed by neutrino telescopes, such as IceCube, Baikal-GVD and KM3NeT. The obtained results can be also applied for description of supernovae neutrino oscillations effects would be detected by JUNO and Hyper-Kamiokande.

This study was conducted within the Scientific Program of the National Center for Physics and Mathematics, Section No. 8 (Stage 2023-2025).
%\section{High-energy neutrino flavour composition}

\end{document}